\documentclass[onecolumn,floatfix,,aps,nofootinbib, showpacs, showkeys]{revtex4}

\usepackage[dvips]{epsfig}
\usepackage[english]{babel}
\usepackage{bbm}
\usepackage{verbatim}
\usepackage{array}
\usepackage{amsmath}
\usepackage{multirow}
\usepackage{amsfonts}
\usepackage{amssymb}
\usepackage{hyperref}
\usepackage{leading}
\usepackage[dvipsnames]{xcolor}
\hypersetup{colorlinks=true,linkbordercolor=Blue,linkcolor=Blue, citecolor=Blue}
\usepackage{subeqnarray}
\usepackage{setspace}
\usepackage{placeins}
\usepackage{float}
\usepackage{indentfirst} 

\usepackage{gensymb}

\begin{document}

\title{Constraints on mapping the Lounesto's Classes}

\author{R. J. Bueno Rogerio$^{1}$} \email{rodolforogerio@unifei.edu.br}

\affiliation{$^{1}$Instituto de F\'isica e Qu\'imica, Universidade Federal de Itajub\'a - IFQ/UNIFEI, \\
Av. BPS 1303, CEP 37500-903, Itajub\'a - MG, Brazil.}


\begin{abstract}
{\textbf{Abstract.}} The so-called Lounesto's classification engenders six distinct classes of spinors, divided into two sectors: one composed by regular spinors (single-helicity spinors) and the other composed by singular spinors (comprising dual-helicity spinors). In the present essay we develop a mechanism to fully define the right class within the Lounesto's classification a spinor belongs to, without necessity to evaluate the 16 bilinear forms. The analysis lies in the following criteria: a judicious inspection of the phases factor present in both spinor's components. Thus, the machinery developed here works for both regular and singular spinors. 
Taking advantage of the present algorithm, we analyse, under certain conditions, the possibility to transmute between the six classes.    
\end{abstract}


\maketitle

\section{Introduction}\label{intro}
Spinors comprises of mathematical and also physical objects, responsible to describe matter, and may be defined in several different fashions. Within the context of the Clifford's algebra, the spinors are defined to be elements of a left minimal ideal, whereas in the context of group theory we say that the spinors are carriers of the fundamental representation of the group \cite{carmeli,lounestolivro}. The first fundamental aspect related with spinors is that a single given spinor cannot be experienced or even detected. Surely, by the Lorentz group compositions laws related to spinors do not recover the original state when a $2\pi$ rotation are performed on a single spinor. Hereupon, up to our knowledge, a given single spinor cannot be/or represent a physical observable. Nevertheless, even the composite of given mathematical quantities are directly associated to physical observables, thus, the importance of the bilinear covariants (bilinear forms). A algorithm to right define the bilinear forms, thus, recast into define an appropriate dual structure for a given spinor and, then, compute all the bilinear covariants associated to them. 

The well-known Lounesto’s classification \cite{lounestolivro} is a comprehensive and exhaustive algorithm that based on the bilinears covariants associated to a given spinor, embraces the possibility of a vast diversity of spinors. Such classification is divided into two sectors: one comprises of regular and the other singular spinors. 
The Lounesto classification is categorized by six distinct classes and there is nothing that preclude that each class have subclasses. The possibility of finding more classes has not been ruled out in \cite{jotanulo}, however, to the best of our knowledge, the physical interpretation of the spinors that may occupy these new classes is still quite complicated, at least so far, no entity harbouring these classes was, in fact, found or even defined. However, mathematically this was shown to be valid as an extension of Lounesto's classification.

The path followed by Lounesto is to classify spinors according with the physical information encoded in the spinors --- not alluding to dynamic, helicity or any other attribute. In this sense, it is possible to accommodate in the aforementioned classification the Dirac, Majorana and Weyl spinors. The regular classes encompass the Dirac spinors and also single-helicity spinors, which do notnecessarily satisfy the Dirac's dynamic, whereas singular classes are conditioned to embrace flag-dipole spinors \cite{dualtipo4,chengflagdipole}, Majorana (flag-pole) spinors and Weyl spinors, filling in classes 4, 5 and 6 respectively. As we shall see in the scope of the present work, flag-dipole and flag-pole spinors are known to carry a very similar structure, the main difference between them lies on the phases parameters. A curious fact that should be mentioned is related to the Weyl's spinors (class 6), due to its peculiarities \cite{cavalcanticlassification}, it is possible to map any class into class 6, however, the opposite is not true. The Weyl spinors are taken to belong to a univocal class within the Lounesto's classification. The whole path to be followed in this manuscript is based on the canonical form, we not deal with the second quantization (quantum fields) and will not allude to them either. 

The paper is organized as it follows: In the next section we set the spinors notation and introduce the Lounesto's classification. In Sect. \ref{part1} we define a general single-helicity spinor and analyse all the possible phases combinations, displayed in a table, making explicit the restrictions to a spinor to belong to a determined class. Taking advantage of the previous results, in Sect. \ref{part2}, we develop an analogous protocol, however, now for dual-helicity spinors. Thus, in sect. \ref{connectingsectors} we look towards to stablish a link between both sectors of the Lounesto's classification besides analyse the possibility to map the classes. Finally, we conclude.

\section{Elementary review}\label{prelude}
This section is reserved for a brief review on the introductory elements that are necessary for the study carried out in the scope of this paper.

\subsection{Notation, spinors and spinorial components}\label{conceitos}
To obtain an explicit form of a given $\psi(p^{\mu})$ spinor we first call out attention for the rest spinors, $\psi(k^{\mu})$. For an arbitrary momentum $(p^{\mu})$, we have the following condition
\begin{equation}\label{1}
\psi(p^{\mu}) = e^{i\kappa.\varphi}\psi(k^{\mu}),
\end{equation}
where the $\psi(k^{\mu})$ rest frame spinor is a direct sum of the $(1/2,0)$ and $(0,1/2)$ Weyl spinor, which usually is defined as 
\begin{equation}
\psi(k^{\mu}) = \left(\begin{array}{c}
\phi_R(k^{\mu}) \\ 
\phi_L(k^{\mu})
\end{array} \right),
\end{equation}
note that  we define the $k^{\mu}$ rest frame momentum as 
\begin{equation}
k^{\mu}\stackrel{def}{=}\bigg(m,\; \lim_{p\rightarrow 0}\frac{\boldsymbol{p}}{p}\bigg), \; p=|\boldsymbol{p}|,
\end{equation}
moreover, the general four-momentum (in spherical coordinates) 
\begin{equation}\label{momentoesferico}
p^{\mu}=(E, p\sin\theta\cos\phi, p\sin\theta\sin\phi, p\cos\theta).
\end{equation}
Thus, the boost operator is defined as follows
\begin{eqnarray}\label{boostoperator}
e^{i\kappa.\varphi} = \sqrt{\frac{E+m}{2m}}\left(\begin{array}{cc}
\mathbbm{1}+\frac{\vec{\sigma}.\hat{p}}{E+m} & 0 \\ 
0 & \mathbbm{1}-\frac{\vec{\sigma}.\hat{p}}{E+m}
\end{array} \right),
\end{eqnarray}
this yields $\cosh\varphi = E/m, \sinh\varphi=p/m$ with $\hat{\boldsymbol{\varphi}} = \hat{\boldsymbol{p}}$.
Thus, such momentum parametrization allow us to defined the right-hand and left-hand components, in the rest-frame referential, under inspection of the helicity operator it directly provide
\begin{equation}\label{operadorhelicidade}
\vec{\sigma}\cdot\hat{p}\; \phi^{\pm}(k^{\mu}) = \pm \phi^{\pm}(k^{\mu}).
\end{equation}
where $\hat{p}$ stands for the spacial components of the parametrization in \eqref{momentoesferico} and $\sigma$ stands for the Pauli matrix. Thus, the positive helicity component is given by 
\begin{equation}\label{comp-mais}
\phi^{+}(k^{\mu}) = \sqrt{m}e^{ i\vartheta_{1}}\left(\begin{array}{c}
\cos(\theta/2)e^{-i\phi/2} \\ 
\sin(\theta/2)e^{i\phi/2}
\end{array}\right), 
\end{equation} 
and the negative helicity reads\footnote{We reserve the right to omit the right-hand or left-hand component label, as commonly presented in the textbooks, due to the fact that this title comes from the way such components are transformed by Lorentz transformations.}
\begin{equation}\label{comp-menos}
\phi^{-}(k^{\mu}) = \sqrt{m}e^{ i\vartheta_{2}}\left(\begin{array}{c}
\sin(\theta/2)e^{-i\phi/2} \\ 
-\cos(\theta/2)e^{i\phi/2}
\end{array}\right).
\end{equation} 
Such components may be related to each other if one make use of the Wigner's time-reversal operator
\begin{equation}\label{thetawigner}
\Theta = \left(\begin{array}{cc}
0 & -1 \\ 
1 & \;\; 0
\end{array} \right),
\end{equation}
given operator holds the following properties $\Theta^2 = -\mathbbm{1}$ and $\Theta^{-1}=-\Theta$. As firstly noticed on Chapter 1 of Ref \cite{ramond}, it allows one to write
\begin{equation}\label{compsviatheta}
\Theta\phi^{*\;+}(k^{\mu}) = \phi^{-}(k^{\mu}), \;\;  \Theta\phi^{*\;-}(k^{\mu}) = -\phi^{+}(k^{\mu}).
\end{equation}

As remarked in Ref \cite{mdobook}, the presence of the factor $\vartheta$ in \eqref{comp-mais} and \eqref{comp-menos} becomes necessary to set up the framework of eigenpinors of parity or charge conjugation operators or eigenspinors of parity operator. One can verify that under a rotation by an angle $\vartheta$ the Dirac spinors pick up a global sign $e^{\pm i\vartheta/2}$, depending on the related helicity. However, this only happens for eigenspinors of parity operator. For the eigenspinors of charge conjugation operator, the phases factor must be $\vartheta_1=0$ and $\vartheta_2=\pi$ \cite{aaca}. Thus, this judicious combination of $\vartheta_1$ and $\vartheta_2$ ensure locality for field - further details can be found in \cite{mdobook}.

\subsection{The Lounesto's Classification}\label{subLounesto}

Let $\psi$ be an arbitrary spinor field, belonging to a section of the vector bundle $\mathbf{P}_{Spin^{e}_{1,3}}(\mathcal{M})\times\, _{\rho}\mathbb{C}^4$, where $\rho$ stands for the entire representation space $D^{(1/2,0)}\oplus D^{(0,1/2)}$. The usual bilinear covariants associated to $\psi$ reads 
\begin{eqnarray}
\label{Azao} \sigma & = &\psi^{\dag}\gamma_0\psi, \;\mbox{(scalar)}\\  
\label{Bzao}\omega & = & i\psi^{\dag}\gamma_0\gamma_5\psi, \;\mbox{(pseudo-scalar)} \\ 
\mathbf{J}&=&J_\mu \theta^\mu = \psi^{\dag}\gamma_0 \gamma_\mu \psi \theta^\mu, \;\mbox{(vector)} \\
\mathbf{K}&=& K_\mu \theta^\mu = \psi^{\dag}\gamma_0 i \gamma_{0123} \gamma_\mu \psi \theta^\mu, \;\mbox{(axial-vector)}\\ 
\label{Szao} \mathbf{S} &=& S_{\mu \nu} \theta^{\mu \nu} = \frac{1}{2} \psi^{\dag}\gamma_0 i \gamma_{\mu \nu} \psi \theta^\mu \wedge \theta^\nu, \;\mbox{(bi-vector)} 
\end{eqnarray} 
where $\gamma_{0123}:=\gamma_5=-i\gamma_0\gamma_1\gamma_2\gamma_3$ and $\gamma_{\mu\nu} : = \gamma_{\mu}\gamma_{\nu}$. Denoting by $\eta_{\mu \nu}$ the Minkowski metric, the set $\{\mathbbm{1},\gamma
_{I}\}$ (where $I\in\{\mu, \mu\nu, \mu\nu\rho, {5}\}$ is a composed index) is a basis for the Minkowski spacetime
${\cal{M}}(4,\mathbb{C})$ satisfying  $\gamma_{\mu }\gamma _{\nu
}+\gamma _{\nu }\gamma_{\mu }=2\eta_{\mu \nu }\mathbbm{1}$, and $\bar{\psi}=\psi^{\dagger}\gamma_{0}$ stands for the adjoint spinor with respect to the Dirac dual. Yet, the elements $\{ \theta^\mu \}$ are the dual basis of a given inertial frame $\{ \textbf{e}_\mu \} = \left\{ \frac{\partial}{\partial x^\mu} \right\}$, with $\{x^\mu\}$ being the global spacetime coordinates. Also, we are denoting $\theta^{\mu \nu} := \theta^\mu \wedge \theta^\nu$.

In the Dirac's theory, the above bilinear covariants are interpreted respectively  as the  mass of the particle ($\sigma$), the pseudo-scalar ($\omega$) relevant for parity-coupling, the current of probability ($\mathbf{J}$), the direction of the electron spin ($\mathbf{K}$), and the probability density of the intrinsic electromagnetic moment ($\mathbf{S}$) associated to the electron. In general grounds, it is always expected
to associate such bilinear structures to physical observables.

The bilinear forms defined in (\ref{Azao})-(\ref{Szao}) obey the so-called Fierz-Pauli-Kofink (FPK) identities, given by \cite{baylis}
\begin{eqnarray}\label{fpkidentidades}
\label{6}\boldsymbol{J}^2 & = & \sigma^2+\omega^2, \\
J_{\mu}K_{\nu}-K_{\mu}J_{\nu} & = & -\omega S_{\mu\nu} - \frac{\sigma}{2}\epsilon_{\mu\nu\alpha\beta}S^{\alpha\beta}, 
\\
J_{\mu}K^{\mu} & = & 0, \\ 
\label{9}\boldsymbol{J}^2 & = & -\boldsymbol{K}^2.
\end{eqnarray}
where we have used the very definition of the dual basis, $\theta^\mu(\textbf{e}_\nu)=\delta^\mu_\nu$, and similarly $\boldsymbol{K}^2 = K_\mu K^\mu$, both clearly being scalars.

 So, the algebraic constraints presented in \eqref{Azao}-\eqref{Szao} reduce the possibilities of (only) six different spinor classes  (for which $\boldsymbol{J}$ is always non-null\footnote{Interesting enough, by construction, the vector $\boldsymbol{J}$ is always non-null due to the fact that at least one of its components depends exclusively of the sum of the square modulo of the phases factor --- regardless of whether the class is regular or singular. Nonetheless, the same can not be asserted for the axial vector $\boldsymbol{K}$.}), known as \emph{Lounesto's Classification} \cite{lounestolivro}:
\begin{enumerate}
  \item $\sigma\neq0$, $\quad \omega\neq0$;
  \item $\sigma\neq0$, $\quad \omega=0$;
  \item $\sigma=0$, $\quad \omega\neq0$;
  \item $\sigma=0=\omega,$ \hspace{0.5cm} $\textbf{K}\neq0,$ $\quad\textbf{S}\neq0$;
  \item $\sigma=0=\omega,$ \hspace{0.5cm} $\textbf{K}=0,$ $\quad\textbf{S}\neq0$;
  \item $\sigma=0=\omega,$ \hspace{0.5cm} $\textbf{K}\neq0,$ $\quad\textbf{S}=0$,
\end{enumerate}
with classes 1, 2 and 3 satisfying $\textbf{K},\textbf{S} \neq 0$. The spinors belonging to the first three classes are called regular spinors while classes 4, 5 and 6 are labelled as singular spinors  \cite{jcap,Benn}. Spinors describing fermions in field theory are called Dirac spinors, and they may belong to classes 1, 2 or 3, i.e., all Dirac spinors are necessarily regular ones.

As it was shown in \cite{bilineares}, due to the contrasting adjoint structure of the Elko spinors \cite{aaca}, it is extremely necessary to rethink the physical interpretation carried by Lounesto's classification. Quite recently a more general treatment and interpretation for the bilinear forms carrying an adjoint structure, which differs from the Dirac's one, was performed \cite{beyondlounesto}. Such analyse takes into account both charged and uncharged particles, in this vein, it is possible to fully understand the bilinear forms as it follows: $\sigma$ still standing for the invariant length. The four-vector $\textbf{J}$ represents the electric current density for charged particles, whereas for neutral particles it may be understood as the effective electromagnetic current four-vector \cite{giuntineutrino}, the bilinear form $\textbf{K}$ shall be related with the spin-alignment due to a coupling with matter- or electromagnetic-field. And finally, the bi-vector $\textbf{S}$ is related to the electromagnetic momentum density for charged particles, for neutral particles case it may correspond to the spin-density momentum or may represent spin-precession (spin-oscillation) in the presence of a matter or a electromagnetic field \cite[and references therein]{studenikinneutrino2,grigoneutrino,ahluwalianeutrino}. The interpretation of the bilinear $\omega$ is less clear, however, when combined into the FPK identity \eqref{6}, it can be interpreted as probability density for regular spinors \cite{bonorapandora,fabbrinonabelian,roldaofabbri,juliounfolding,lounestolivro}, also some authors claim that such amount gives us clues of spinor behavior under $CPT$ symmetry \cite{lounestolivro}. Nonetheless, for the general classification as developed in \cite{beyondlounesto}, the physical interpretation of the bilinear $\omega$ is not clear.

\section{Part 1: Further investigations on the regular spinors}\label{part1}

Consider a single-helicity spinor, which can be described in the Weyl representation as follows
 \begin{eqnarray}
\psi &=& \left(\begin{array}{c}
\alpha\phi_R\\
\beta\phi_L
\end{array}
\right),
\end{eqnarray}
where in its algebraic form reads
\begin{eqnarray}\label{psi}
\psi=\left(\begin{array}{c}
\alpha a\\
\alpha b\\
\beta c\\
\beta d
\end{array}
\right),
\end{eqnarray}
where $a, b, c, d \in \mathbbm{C}$ and the phases factors $\alpha$ and $\beta$ $\in\mathbbm{C}$ which will be further determined. The only requirement under the phases factors, comes from the orthonormal relation (invariant norm), and it stands for $|\alpha|^2+|\beta|^2\propto m$, in which $m$ stands for the mass of a particle, being a real number.

Up two possibilities, first we impose to the spinor \eqref{psi} to carry a positive single helicity, where we have used the plus sign in Eq. \eqref{operadorhelicidade}, thus, we obtain the following relations \cite{interplay}
\begin{eqnarray}\label{b+}
b=\frac{a\sin\theta}{1+\cos\theta}e^{\textit{i}\phi} \;\;\;\; \mbox{and} \;\;\;\; d=\frac{c\sin\theta}{1+\cos\theta}e^{\textit{i}\phi}.
\end{eqnarray}
Under the conditions above, we can write the positive helicity spinor as 
\begin{eqnarray}\label{positive}
\psi_{(+,+)}=\left(\begin{array}{c}
\alpha a\\
\alpha\frac{a\sin\theta}{1+\cos\theta}e^{i\phi}\\
\beta c\\
\beta\frac{c\sin\theta}{1+\cos\theta}e^{i\phi}
\end{array}
\right),
\end{eqnarray}
where the lower index stand for helicity of the right-hand and left-hand components, respectively. Analogously, for a negative single helicity spinor, taking into account the negative sign in Eq.\eqref{operadorhelicidade}, we define
\begin{eqnarray}\label{b-}
b=-\frac{a\sin\theta}{1-\cos\theta}e^{\textit{i}\phi} \;\;\;\; \mbox{and} \;\;\;\; d=-\frac{c\sin\theta}{1-\cos\theta} e^{i\phi},
\end{eqnarray}
and then, the negative helicity spinor reads
\begin{eqnarray}\label{shp}
\psi_{(-,-)}=\left(\begin{array}{c}
\alpha a\\
-\alpha\frac{a\sin\theta}{1-\cos\theta} e^{i\phi}\\
\beta c\\
-\beta\frac{c\sin\theta}{1-\cos\theta} e^{i\phi}
\end{array}
\right).
\end{eqnarray}
A parenthetic remark, the above single-helicity spinors do not necessarily fulfil the Dirac dynamics. Once imposed the Dirac dynamics automatically parity operation must play the central role connecting the representation spaces \cite{ryder,speranca,diracpauli}. However, here, for the proposal of the paper, taking into account that we are in an abstract framework, so, we do not make assertions related to the dynamic.
With the relations \eqref{comp-mais} and \eqref{comp-menos} at hands, one may write the introduced single-helicity spinors in the following fashion\footnote{In Eqs.\eqref{psisingle} and \eqref{psidual} we freely omitted (or absorbed in the phases factor) the Lorentz boost factors. Since the boost operators have already acted on the spinor's components providing only a constant factor, now what really matters to us is just the helicity carried by spinor.}
\begin{equation}\label{psisingle}
\psi_{(+,+)} = \sqrt{m}\left(\begin{array}{c}
\alpha\cos(\theta/2)e^{-i\phi/2} \\ 
\alpha\sin(\theta/2)e^{i\phi/2} \\ 
\beta\cos(\theta/2)e^{-i\phi/2} \\ 
\beta\sin(\theta/2)e^{i\phi/2}
\end{array}\right),\; \psi_{(-,-)} = \sqrt{m}\left(\begin{array}{c}
-\alpha\sin(\theta/2)e^{-i\phi/2} \\ 
\alpha\cos(\theta/2)e^{i\phi/2} \\ 
-\beta\sin(\theta/2)e^{-i\phi/2} \\ 
\beta\cos(\theta/2)e^{i\phi/2}
\end{array}\right).
\end{equation} 
Note that to date we have not alluded to which class of regular spinors these spinors belongs to. Now we apply the mathematical proposal to the above single-helicity spinors. The procedure, then, lies in the analysis of the phases factors $\alpha$ and $\beta$, aiming to investigate under what circumstances one is able define the classes that the above spinors belong by an inspection of the restrictions upon its values, note that such a classification procedure does not require us to evaluate the bilinear forms associated with spinor. Thus, the conditions between the phases are displayed in the following table: 
\begin{table}[H]
\centering
\begin{tabular}{c|c|c|c}
\hline
\multicolumn{4}{c}{\textbf{Single-helicity spinors}}\\
\hline 
\hline 
 \;\;\;\;\;\;\;\;\;\;$\alpha$\;\;\;\;\;\;\;\;\;\; & \;\;\;\;\;\;\;\;\;\;$\beta$\;\;\;\;\;\;\;\;\;\; & \;\;\;\;\;\;\;\;\;\;Class\;\;\;\;\;\;\;\;\;\; & Constraints \\ 
\hline 
\hline 
 ${\rm I\!R}$ & ${\rm I\!R}$ & 2 & - \\ 
\hline 
$\mathbb{C}$ & $\mathbb{C}$ & 2 & $\alpha=\beta$ \\ 
\hline 
 $\mathbb{C}$ & $\mathbb{C}$ & 1 & - \\ 
\hline 
 $\mathbb{C}$ & ${\rm I\!R}$ & 1 & - \\ 
\hline 
 $\mathbb{C}$ & ${\rm Im}$ & 1 & - \\ 
\hline 
 ${\rm Im}$ & ${\rm Im}$ & 2 & - \\ 
\hline 
 ${\rm Im}$ & ${\rm I\!R}$ & 3 & - \\ 
\hline 
 0 & ${\rm I\!R}$, $\mathbb{C}$ or ${\rm Im}$ & 6 & - \\ 
\hline 
 ${\rm I\!R}$, $\mathbb{C}$ or ${\rm Im}$  & 0 & 6 & - \\ 
\hline 
\hline 
\end{tabular}
\caption{The phases constraints to classify regular spinors.}
\end{table} 
\noindent where ${\rm I\!R}$ denotes any real number, $\mathbb{C}$ stands for any c-number, ${\rm Im}$ stands for any purely imaginary number and the symbol ``-'' means ``No Constraint''. 
Imposing to \eqref{psisingle} to satisfy the Dirac's equation, automatically the link that arises between the phases is $\alpha=\beta$, thus, belonging to class 2.

As an example of the utility of the protocol above, it is possible to correctly classify the single-helicity spinors introduced \cite{diracpauli,nondirac,dinorestricted}. 
Note that we covered all the regular classes besides class 6. Remarkably enough, as one can see in Ref \cite{cavalcanticlassification}, class 6 spinors necessarily have or $\phi_R=0$ or $\phi_L=0$, evincing, thus, the impossibility (independent of the choice of the phases) to map class 6 into any other class.    
We emphasize, however, that a single-helicity spinor can never be regarded as a singular spinor, note that, for the algebraic spinor displayed in \eqref{psi}, the computation of the bilinears $\sigma$ and $\omega$ provides 
\begin{equation}\label{sigma}
\sigma = \alpha\beta^*+\alpha^*\beta,
\end{equation}
and 
\begin{equation}\label{omega}
\omega = i(\alpha\beta^*-\alpha^*\beta),
\end{equation}
if one impose to \eqref{sigma} and \eqref{omega} to be simultaneously null, a contradiction is automatically reached. Note that 
\begin{equation}
\sigma = 0 \;\;\longrightarrow\;\; \alpha\beta^*=-\alpha^*\beta, 
\end{equation}
and 
\begin{equation}
\omega = 0 \;\;\longrightarrow\;\; \alpha\beta^*=\alpha^*\beta, 
\end{equation}
the only way to satisfy both condition is, then, $\alpha=0$ or $\beta=0$, trivially leading to class 6 (excluding all the other classes), and the other possibility stands for the phases product to be zero, i.e., $\alpha\beta^*=0=\alpha^*\beta$, showing, then, an inconsistency and  also a non-physical case. Thus, the arguments above reinforce that single-helicity spinors can never fit into singular classes.

\section{Part 2: Further investigations on the singular spinors}\label{part2}

In this section, we look towards define dual-helicity spinors. Such a task is accomplished by using a very similar procedure as previously performed for singular-helicity spinors. Now, mixing the signs in Eq.\eqref{operadorhelicidade}, then, we have the following conditions $\vec{\sigma}\cdot \hat{p}\; \phi_{R}^{+}= +\phi_{R}^{+}$ combined with $\vec{\sigma}\cdot \hat{p}\; \phi_{L}^{-}= -\phi_{L}^{-}$, as well as $\vec{\sigma}\cdot \hat{p}\; \phi_{R}^{-}= -\phi_{R}^{-}$ with $\vec{\sigma}\cdot \hat{p}\; \phi_{L}^{+}= +\phi_{L}^{+}$. The requirements above translates into the following set dual-helicity spinors, 
\begin{eqnarray}\label{dual1}
\psi_{(+,-)}=\left(\begin{array}{c}
\alpha a\\
\alpha\frac{a\sin\theta}{1+\cos\theta} e^{i\phi}\\
\beta c\\
-\beta\frac{c\sin\theta}{1-\cos\theta} e^{i\phi}
\end{array}
\right), \;
\psi_{(-,+)}=\left(\begin{array}{c}
\alpha a\\
-\alpha \frac{a\sin\theta}{1-\cos\theta} e^{i\phi}\\
\beta c\\
\beta \frac{c\sin\theta}{1+\cos\theta} e^{i\phi}
\end{array}
\right),
\end{eqnarray}
or in a compact form, 
\begin{eqnarray}\label{dual2}
\psi_{(\pm,\mp)}=\left(\begin{array}{c}
-\alpha\frac{bcd^{*}}{|c|^{2}}\\
\alpha b\\
\beta c\\
\beta d
\end{array}
\right).
\end{eqnarray}
The spinors in \eqref{dual1} can be written as
\begin{equation}\label{psidual}
\psi_{(+,-)}=\sqrt{m}\left(\begin{array}{c}
\alpha\cos(\theta/2)e^{-i\phi/2}\\
\alpha\sin(\theta/2)e^{i\phi/2}\\
-\beta\sin(\theta/2)e^{-i\phi/2}\\
\beta\cos(\theta/2)e^{i\phi/2}\\
\end{array}
\right), \;\;\psi_{(-,+)}=\sqrt{m}\left(\begin{array}{c}
-\alpha\sin(\theta/2)e^{-i\phi/2}\\
\alpha\cos(\theta/2)e^{i\phi/2}\\
\beta\cos(\theta/2)e^{-i\phi/2}\\
\beta\sin(\theta/2)e^{i\phi/2}
\end{array}
\right).
\end{equation}
An important remark lies in the fact that dual spinors components are always connected via Wigner's time-reversal operator ($\Theta$), similar to Elko spinors \cite{mdobook} and Flag-dipole spinors \cite{dualtipo4}. 
Now we display the conditions satisfied between the phases to classify dual-helicity spinors within Lounesto's classification:

\begin{table}[H]
\centering
\begin{tabular}{c|c|c|c}
\hline
\multicolumn{4}{c}{\textbf{Dual-helicity spinors}}\\
\hline 
\;\;\;\;\;\;\;\;\;\;$\alpha$\;\;\;\;\;\;\;\;\;\; & \;\;\;\;\;\;\;\;\;\;$\beta$\;\;\;\;\;\;\;\;\;\; & \;\;\;\;\;\;\;\;\;\;Class\;\;\;\;\;\;\;\;\;\; & \;\;\;Constraints\;\;\; \\ 
\hline 
\hline 
${\rm I\!R}$ & ${\rm I\!R}$ & 4 & $\alpha\neq\beta$ \\ 
\hline 
${\rm I\!R}$ & ${\rm I\!R}$ & 5 & $\alpha=\beta$ \\ 
\hline 
$\mathbb{C}$ & $\mathbb{C}$ & 4 & $|\alpha|^2\neq|\beta|^2$ \\
\hline
 $\mathbb{C}$ & $\mathbb{C}$ & 5 & $|\alpha|^2=|\beta|^2$ \\
\hline 
$\mathbb{C}$ & ${\rm I\!R}$ & 4 & $|\alpha|^2\neq|\beta|^2$ \\ 
\hline 
$\mathbb{C}$ & ${\rm I\!R}$ & 5 & $|\alpha|^2=|\beta|^2$ \\ 
\hline 
$\mathbb{C}$ & ${\rm Im}$ & 4 & $|\alpha|^2\neq|\beta|^2$ \\ 
\hline
$\mathbb{C}$ & ${\rm Im}$ & 5 & $|\alpha|^2=|\beta|^2$ \\ 
\hline
${\rm Im}$ & ${\rm Im}$ & 5 & $|\alpha|^2=|\beta|^2$  \\ 
\hline 
${\rm Im}$ & ${\rm Im}$ & 4 & $|\alpha|^2\neq|\beta|^2$ \\ 
\hline 
${\rm Im}$ & ${\rm I\!R}$ & 4 & $|\alpha|^2\neq|\beta|^2$ \\ 
\hline 
${\rm Im}$ & ${\rm I\!R}$ & 5 & $|\alpha|^2=|\beta|^2$ \\ 
\hline
0 & ${\rm I\!R}$, $\mathbb{C}$ or ${\rm Im}$ & 6 & - \\ 
\hline 
${\rm I\!R}$, $\mathbb{C}$ or ${\rm Im}$  & 0 & 6 & - \\ 
\hline 
\hline 
\end{tabular} 
\caption{The phases constraints to classify singular spinors.}
\end{table} 
\noindent Thus, we turn explicit the constraints among the phases to define a specific class. Regarding to the class 4 and 5, we lay emphasis on their difference. The distinction among the aforementioned spinors is present on the value of the bilinear form $\boldsymbol{K}$. Such bilinear form, in general grounds, depends on the subtraction between the modulo of the phases, thus, if $|\alpha|^2=|\beta|^2$ it leads to a spinor to belong to class 5. On the other hand, for a spinor carrying $|\alpha|^2\neq|\beta|^2$, it provides $\boldsymbol{K}\neq 0$, leading, then, to belong to class 4, as it can easily been checked for the spinors in Refs\cite{tipo4mdo,dualtipo4,chengflagdipole}.  

We highlight an important fact here, dual-helicity spinors, independent of the phases, are automatically singular spinors and can not never be classified as regular spinors.

\section{Connecting the Lounesto's Regular and Singular sectors}\label{connectingsectors}
So far we have shown how to identify the class of a given regular or singular spinor just by the phase analysis, without invoking the computation of the bilinear forms. The next step, is to show how to connect both sectors of the Lounesto's classification 
\begin{eqnarray}
&&\left. \begin{array}{rlll}
1. & \sigma\neq0, & \omega\neq0,  \\
2. & \sigma\neq0, &\omega=0,  \\
3. & \sigma=0, & \omega\neq0, 
\end{array}\right\}\mbox{\emph{Single-helicity sector}}. 
\nonumber\\
&&\left. \begin{array}{rlll}
4. & \sigma=0=\omega, & \boldsymbol{K}\neq0, & \boldsymbol{S}\neq0, \\
5. & \sigma=0=\omega, & \boldsymbol{K}=0, & \boldsymbol{S}\neq0,  \\
\end{array}\right\}\mbox{\emph{Dual-helicity sector}}.
\\\nonumber
&&\left. \begin{array}{rlll}
6. & \sigma=0=\omega, & \boldsymbol{K}\neq0, & \boldsymbol{S}=0.
\end{array}\right\}\mbox{\emph{Not defined}}.
\end{eqnarray} 

Given what has already been discussed in the previous sections, we will abstain from the analysis of the class 6 for obvious reasons. Thus our focus lies in the classes comprised between 1 to 5. Have seen, that Lounesto's encompasses single- and dual-helicity spinors, the only way to connect both sectors is transmuting the spinor's helicity. Such task is easily accomplished via Wigner's time-reversal operator and algebraic complex conjugation, as shown in Eq. \eqref{compsviatheta}. Taking into account such peculiarities, one is able, then, to (re)define the phases as it reads
\begin{equation}\label{alfalinha}
\alpha\rightarrow\alpha^{\prime}=\alpha\hat{\Delta},
\end{equation}   
and 
\begin{equation}\label{betalinha}
\beta\rightarrow\beta^{\prime}=\beta\Theta\hat{\Delta},
\end{equation} 
where we have defined $\hat{\Delta}\equiv\Theta\mathcal{K}$, in which $\mathcal{K}$ stands for the algebraic complex-conjugation and $\Theta$ is defined in \eqref{thetawigner}. Such a redefinition above, allow one to define a more involving operator, which act on spinors. Such operator read
\begin{equation}\label{Sigmao}
\Sigma_{\alpha} = \left(\begin{array}{cc}
\alpha^{\prime} & 0 \\ 
0 & \mathbbm{1}
\end{array} \right),  \;\Sigma_{\beta} = \left(\begin{array}{cc}
\mathbbm{1} & 0 \\ 
0 & \beta^{\prime}
\end{array} \right).
\end{equation}
Now, let $\Gamma_{R}= \{\sigma_R, \omega_R, \boldsymbol{J}_R, \boldsymbol{K}_R, \boldsymbol{S}_R\}$ be a set of bilinear forms of a given spinor belonging to the regular sector and $\Gamma_{S}= \{\sigma_S, \omega_S, \boldsymbol{J}_S, \boldsymbol{K}_S, \boldsymbol{S}_S\}$ be a set of bilinear forms of a given spinor belonging to the singular sector, the conditions in \eqref{alfalinha} and \eqref{betalinha} together with \eqref{Sigmao}, ensure the following
\begin{eqnarray*}
\Sigma: \psi_{R} &\mapsto & \psi_{S},
\\
\Gamma_{R} &\mapsto & \Gamma_{S},
\end{eqnarray*}
where the lower index $R$ and $S$ stands for regular and singular sectors. Obviously, the inverse procedure indeed exists.
Note that, however, to map a single-helicity spinor into a dual-helicity spinor, one may make use of only one of the two relations \eqref{alfalinha} and \eqref{betalinha}.

Now, taking advantage of \eqref{Sigmao}, if one act on the spinor introduced in \eqref{psisingle} easily obtain the same relations presented in \eqref{psidual}, connecting the Lounesto's regular sector with the singular sector. Based on the two previous tables, one is able to define the following 
\begin{table}[H]
\centering
\begin{tabular}{c|c|c|c|c|c|c}
\hline 
 \multicolumn{3}{c|}{\textbf{Single-helicity spinors}} &    \multicolumn{3}{c|}{\textbf{Dual-helicity spinors}} &   \\ 
\hline 
\hline 
\;\;\;\;\;\;\;\;$\alpha$\;\;\;\;\;\;\;\; &\;\;\;\;\;\;\;\; $\beta$\;\;\;\;\;\;\;\; &\;\;\;\;\;\;\;\; Class\;\;\;\;\;\;\;\; & \;\;\;\;\;\;\;\;$\alpha$\;\;\;\;\;\;\;\;& \;\;\;\;\;\;\;\;$\beta$\;\;\;\;\;\;\;\; & \;\;\;\;\;\;\;\;Class\;\;\;\;\;\;\;\; & Constraints \\ 
\hline 
${\rm I\!R}$ & ${\rm I\!R}$ & 2 & ${\rm I\!R}$ & ${\rm I\!R}$ & 4 (5) & $\alpha\neq\beta$ ($\alpha=\beta$) \\ 
\hline 
$\mathbb{C}$ & $\mathbb{C}$ & 2 & $\mathbb{C}$ & $\mathbb{C}$ & 5 & $\alpha=\beta$  \\ 
\hline 
$\mathbb{C}$ & $\mathbb{C}$ & 1 & $\mathbb{C}$ & $\mathbb{C}$ & 4 & $\alpha\neq\beta$ \\ 
\hline 
$\mathbb{C}$ & $\mathbb{C}$ & 1 & $\mathbb{C}$ & $\mathbb{C}$ & 5 & $|\alpha|^2=|\beta|^2$ \\ 
\hline 
$\mathbb{C}$ & ${\rm I\!R}$ & 1 & $\mathbb{C}$ & ${\rm I\!R}$ & 4 (5) & $|\alpha|^2\neq|\beta|^2$ ($|\alpha|^2=|\beta|^2$) \\ 
\hline 
$\mathbb{C}$ & ${\rm Im}$ & 1 & $\mathbb{C}$ & ${\rm Im}$ & 4 (5) & $|\alpha|^2\neq|\beta|^2$ ($|\alpha|^2=|\beta|^2$) \\ 
\hline 
${\rm Im}$ & ${\rm Im}$ & 2 & ${\rm Im}$ & ${\rm Im}$ & 4 (5) & $|\alpha|^2\neq|\beta|^2$ ($|\alpha|^2=|\beta|^2$) \\ 
\hline 
${\rm Im}$ & ${\rm I\!R}$ & 3 & ${\rm Im}$ & ${\rm I\!R}$ & 4 (5) & $|\alpha|^2\neq|\beta|^2$ ($|\alpha|^2=|\beta|^2$) \\ 
\hline 
\hline 
\end{tabular}
\caption{Constraints to map the regular and singular sector of the Lounesto's classification.} 
\end{table}
\noindent Note, however, the possibility to map classes 1, 2, 3 into classes 4 and 5. As we shall see, the protocol developed here is in agreement with the previous one performed in \cite{juliojmp07}, where the authors map Dirac spinors into Elko spinors (and vice-versa), showing the possibility to connect both sectors of the Lounesto's classification. Further investigations, but now concerning the dynamics of Dirac and Elko field are taken into account, were performed in \cite{fromdiractoelko}. Starting with a fundamental action, representing a mass dimension-transmuting operator between Dirac and Elko spinor fields were performed on the Dirac Lagrangian, in order to lead it into the Elko Lagrangian. Such a programme allow Elko spinors  to be incorporated in the Standard Model. Such a possible connection between both sectors show - up to our knowledge - an useful tool to understand the underlying properties of the spinors encompassed into the Lounesto's classification. 

We purposely omitted, but all classes can be mapped into class 6.

\section{Final Remarks}

In the present report we introduced a set of single-helicity and dual-helicity spinors on its algebraic form. After defining the eigenstates of the helicity operator, we explicitly defined the spinors in terms of spherical coordinates. Thus, with such spinors at hands, we developed a mechanism which turns possible to right ascertain to what class the spinors belongs within Lounesto's classification, without an exhaustive computation of the bilinear forms. Such a mathematical device is based on the analyses of the arbitrary phases factors, being them real numbers, complex numbers or purely imaginary numbers, thus, we displayed in two tables all the possible combinations among the phases, one table for regular spinors and the other one for singular spinors. As highlighted above, we emphasize again that Lounesto's classification show a strong dichotomy between the regular sector and the singular sector. 

By a judicious inspection of the structure of the bilinear forms, one can show the impossibility --- given by mathematical inconsistencies --- to a given single-helicity spinor belong to the singular sector and we also evinced that dual-helicity spinors can not belong to the regular sector.

We shall finalize making an allusive comment to the results found in Sect.\ref{connectingsectors}, as we can see, the phases factors play the central role in connecting both sectors of the Lounesto's classification, through the study that was developed here, we can also say that a portion of the physical information carried by the spinors is encoded in the phases factor.

\section{Acknowledgements}
RJBR thanks CNPq Grant N$^{\circ}$. 155675/2018-4 for the financial support.

\bibliographystyle{unsrt}
\bibliography{refs}

\end{document}